\documentclass[conference]{IEEEtran}
\IEEEoverridecommandlockouts
\usepackage{cite}
\usepackage{amsmath,amssymb,amsfonts}
\usepackage{algorithmic}
\usepackage{graphicx}
\usepackage{textcomp}
\usepackage{xcolor}
\usepackage[hidelinks]{hyperref}
\def\BibTeX{{\rm B\kern-.05em{\sc i\kern-.025em b}\kern-.08em
    T\kern-.1667em\lower.7ex\hbox{E}\kern-.125emX}}
    
\usepackage{xstring}
\newcommand{\PP}[1]{
\vspace{4px}
\noindent{\bf \IfEndWith{#1}{.}{#1}{#1.}}
}

\newcommand{\PPi}[1]{
\vspace{0px}
\noindent{\it \IfEndWith{#1}{.}{#1}{#1.}}
}

\begin{document}

\title{Edge Security: Challenges and Issues}

\author{

\IEEEauthorblockN{Xin Jin, Charalampos Katsis, Fan Sang, Jiahao Sun, Ashish Kundu, Ramana Kompella}
\IEEEauthorblockA{
Cisco Research\\
\{xijin3, ckatsis, fsang, jiahasun, ashkundu, rkompell\}@cisco.com}
}





\maketitle
\thispagestyle{plain}
\pagestyle{plain}

\begin{abstract}
Edge computing is a paradigm that shifts data processing services to the network edge, where data are generated. While such an architecture provides faster processing and response, among other benefits, it also raises critical security issues and challenges that must be addressed. 

This paper discusses the security threats and vulnerabilities emerging from the edge network architecture spanning from the hardware layer to the system layer. We further discuss privacy and regulatory compliance challenges in such networks. Finally, we argue the need for a holistic approach to analyze edge network security posture, which must consider knowledge from each layer.
\end{abstract}






\section{Introduction}
An edge computing infrastructure is being adopted widely in the industry for better performance, lower latency, better data governance, privacy, and compliance. With the exponential growth in the usage of Internet-of-Things (IoT) devices and sensors\footnote{IoTs are being used in manufacturing industry (Industry 4.0~\cite{lasi2014industry}), healthcare, personal health management, agriculture, transportation, and across other sectors~\cite{iotww}.}, sending petabytes of data to a centralized cloud or an enterprise data center has several challenges. The processing of IoT data incurs huge delays and costs, and technical challenges to data movement, management, processing, storage, privacy, security, and compliance. As an extension to the cloud and data center perimeter, the edge computing infrastructure deploys edge devices, servers, and edge-network capabilities near (within 1-hop of) data sources. 

The increase in the size of edge computing infrastructure and its deployment in a geo-distributed manner provides better performance and data governance. However, at the same time, it offers new security, privacy, and regulatory compliance challenges. Security of edge computing involves various fields, such as networks, systems, cryptography, and machine learning. Hindered by the diversity of domain-specific knowledge, security researchers specialized in each field are investigating the challenges without a holistic view of the edge computing system. Such an isolated research methodology overlooks security challenges that emerge from the interaction boundaries between domains. Worse, naively fitting a domain-specific security measurement to the overall edge computing pipeline brings incompatibility and redundancy issues. 

In this paper, we explore the emerging security vulnerabilities and threats specific to edge computing spanning across different domains. Our goal is to stress the security issues arising from the different layers and spark initiatives to establish a uniform system that holistically analyzes them.

\section{Background and Overview}
\subsection{Edge Computing Framework}

\begin{figure}[ht]
\centering
\includegraphics[width=0.95\linewidth]{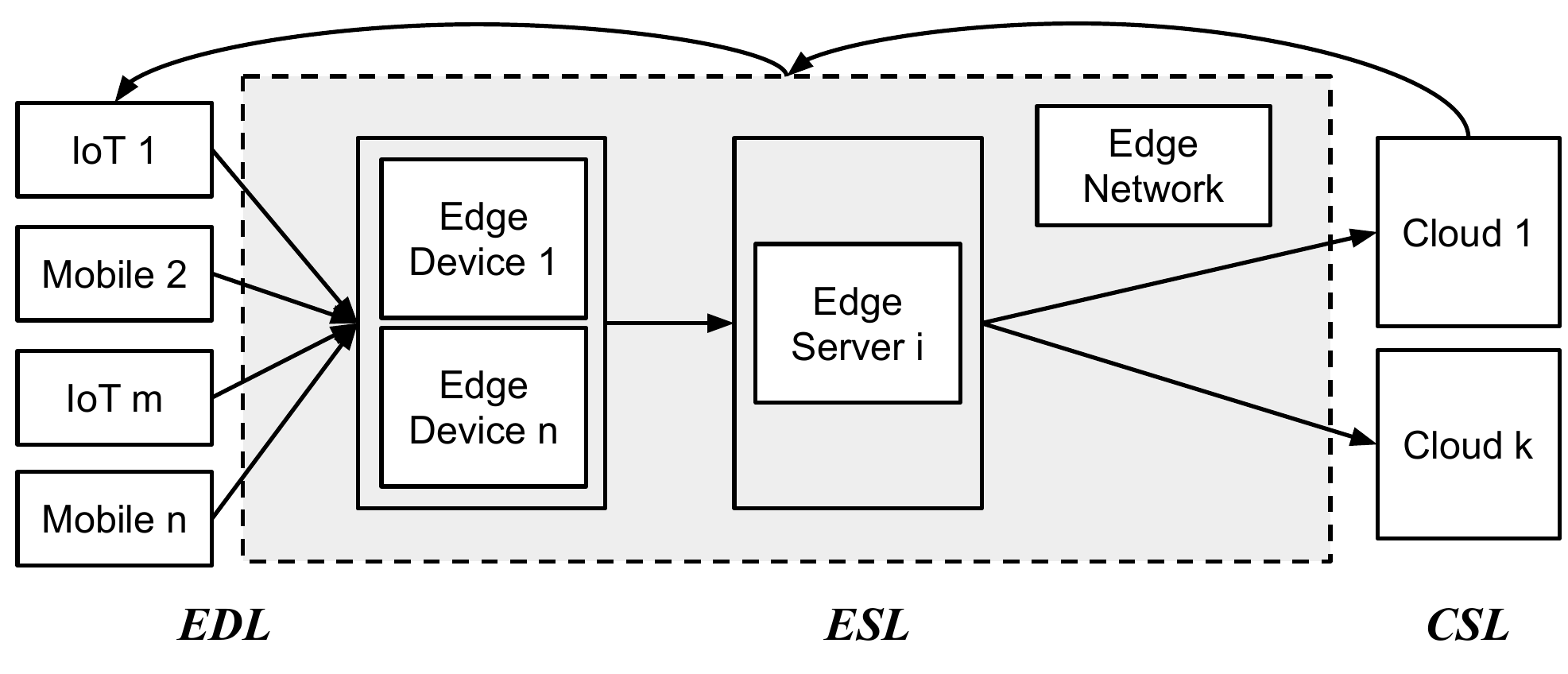}
\vspace{-0.1in}
\caption{The general architecture of edge computing}
\label{fig:edge-arch}
\end{figure}

An edge computing infrastructure
has edge devices as well as edge servers
as primary entities to
carry out computation and data processing.
A general architecture of edge computing
mainly consists of three layers:
an edge device layer (EDL),
an edge server layer (ESL),
and a cloud server layer (CSL)~\cite{edge-sok}.
Systems on the CSL
have the most powerful computational power,
the ones on the ESL follow, and
the devices on the EDL
usually have the least computational power.

\PP{Edge Device Layer (EDL)}
Devices deployed to the EDL
are called \textit{edge devices},
which usually conduct field tasks
such as sensing, actuating, and controlling,
in the physical world.
Common edge devices
are controlled by microcontrollers (MCUs),
which implements the firmware
that provides low-level software interfaces
to control the device’s hardware~\cite{mcu}.
%
%
Edge devices can be further
grouped into
\textit{IoT devices}
and \textit{mobile devices}.
%
IoT devices usually consist of
lightweight electronic devices
that are interconnected or connected
to the ESL through wireless protocols
such as 4G/5G, WiFi, and Bluetooth.
Examples include
smart home devices,
health monitoring devices,
and smart warehouse carts in industrialized IoT (IIoT).
%
%
%
Most IoT devices
use Cortex-M series MCUs produced
by STMicroelectronics~\cite{cortex}.
Once a real-time operating systems (RTOS)~\cite{rtos, rtos2, rtos3}
is burned into the IoT devices,
no additional programming interfaces
are provided under normal circumstances.
Meanwhile,
mobile devices usually have
more advanced operating systems,
such as Android and iOS,
which provide programmable interfaces
for programmers to develop
applications compatible with the OS.
Common mobile devices
include smartphones, tablets,
and central controllers of smart vehicles,
and usually adopt Cortex-A series MCUs produced
by high-performance chip manufacturers such as Qualcomm~\cite{qualcomm}.

\PP{Edge Server Layer (ESL)}
Edge servers handle
the core computing functions
in edge computing,
including authentication,
authorization,
computation,
data analytics,
task offloading,
and data storage.
ESL usually consists of
multiple hierarchical sub-layers of edge servers
with increasing computational power.
Devices such as
wireless base stations and access points (APs)
sit at the lowest sub-layer.
They are mainly responsible for
receiving data from the edge devices and
transmitting control flows back to them.
Base stations/APs forward the data
received from edge devices to
the edge servers in the higher layer
to perform individual computation tasks.
%
%
When a task is
too complicated to be handled
on the current edge server,
the task will be delegated to the servers
with more computational power at a even higher sub-layer.
After the task is handled properly,
a sequence of control flows
will be passed back to the base stations/APs,
and eventually transmitted to the edge devices.
Popular state-of-the-art edge servers include
NVIDIA Jetson Nano~\cite{jetson},
Raspberry Pi~\cite{pi},
Marvell OCTEON 10 DPU~\cite{marvell},
and Mac Mini~\cite{macmini}.

\PP{Cloud Service Layer (CSL)}
CSL hosts cloud servers
and data centers.
The cloud servers
are responsible for the highest level
of operations and
integration of tasks offloaded from EDL and ESL,
while the data centers
storing the vast amount of data
generated in the edge computing infrastructure.
Usually,
CSL consists of clusters of powerful machines.
In this paper,
we mainly focus on EDL and ESL,
as they are unique infrastructures in
edge computing.


\subsection{Security Characteristics of Edge Computing}

Even though edge computing
and cloud computing are similar
with respect to
the offered services and functionalities,
the scopes of security measures are largely different,
posing new challenges due to
the unique characteristics of edge computing platforms.

\PP{Weaker computation power}
Compared to a cloud server,
the computation power of an edge server is
relatively weaker.
As a result,
an edge server might be more vulnerable to
existing attacks that are less effective
towards a cloud server.
Similarly,
compared to general-purpose computers,
edge devices have less robust defense systems.
Attacks that have been
mitigated for desktop computers
can still pose serious threats.

\PP{Large volume of interconnected devices}
Client devices in cloud computing
are usually not interconnected,
limiting the influential impact
when a few are compromised.
Nevertheless,
edge devices are typically interconnected,
so a small intrusion can
have a more significant impact
if the attack is spread and
propagated among devices (e.g,. Mirai botnet~\cite{mirai}).
%

\PP{Heterogeneous device form factors}
As we can see,
devices of various form factors
can co-exist in edge computing,
especially in EDL,
where depending on the purpose of the device
and the physical location being deployed,
two devices might have totally different
hardware and software stacks.
Such a heterogeneity of devices
poses extreme challenges
when designing general solutions
to potential threats and issues.
The number of different scenarios
of devices interacting with each other
increases exponentially when the
number of devices increases.
Handling all those scenarios
either sacrifices generality if target
a specific subgroup of edge devices,
or loses accuracy when searching for a panacea.

\PP{Unavailability of security features}
Various hardware security features have
landed on modern platforms (e.g., \cite{ibm, apple, tpm, smart})
to mitigate existing and unforeseen threats.
%
%
Meanwhile,
architectural improvements
are being introduced to the CPU,
as well as its chipset~\cite{tee, nx, mpx, mpk, tsx, cat}.
%
Not to mention
the collection of software mitigation techniques
built on top of those platform-specific features.
Unfortunately,
due to different form factors and diverse of platforms,
desired security features are not always available
on specific edge computing platforms,
and a tough decision of
security versus cost-effectiveness
has to be made.

\PP{Maintaining quality of service}
Last but not least,
any additional security measurements should try to maintain
the original quality of service (QoC) (e.g., availability and real-timeness)
at best effort,
as it is the initial purpose of edge computing.

\subsection{Multi-dimensional Security Analysis of Edge Computing}
The logical components
of an edge computing stack are:
device hardware, firmware and system,
network and communication,
cloud stack (such as Kubedge~\cite{kubedge}),
machine learning as workload infrastructure and compute apps,
data protection,
privacy,
cryptography,
users,
identity and access management,
and regulatory compliance.

In this paper
we amplify an edge device and dissect
in order to study its security issues.
This study can be extended to
an edge server or
any other cyberphysical edge entity.
In the following sections,
we will systematically study the vulnerabilities
specific to edge computing for each of these logical components.

\section{Device and System} 


As the backbone of edge computing, the edge device itself and the system on top serve as the shield of the edge computing system
at the lowest level. 
System software such as the OS often runs at the highest privilege and mediates the management tasks across process domains. At the same time, the hardware provides the resources for the OS to correctly fulfill tasks as intended by the user.
As a result, the hardware and system software define the trustworthiness of a standalone edge device, which is a fundamental building block to achieving a secure edge computing service and fully unleashing its power.

We define the hardware and the system software of an edge device as an \textit{edge platform}. To support the desired security requirements of edge computing services, an ideal edge platform should consist secure 1) physical protection (\ref{ss:physical}) 2) hardware components (\ref{ss:hw}), 3) firmware (\ref{ss:firmware}), and 4) system software (e.g., OS) (\ref{ss:system}).

\PP{Re-imagining existing threats under edge computing}
As computing devices themselves of different form factors, edge platforms are susceptible to existing security threats if they possess the targeted attack surface.
%
Worse, the unique characteristics of edge computing may exacerbate the impact of such attacks.
Together with privileged attackers that possess and provision the edge devices of diverse form factors, there are various challenges to guarantee a trustworthy edge platform.  

\subsection{Insecure Physical Protection}
\label{ss:physical}
Physical access is commonly assumed as the last layer to defend against
attackers. Generally speaking, physical access is always excluded
from threat models of cloud computing. 
However, this is one of the initial weapons granted in the attacker's arsenal under edge computing. 

\begin{enumerate}
\item \textit{Intrusive attacks} require physical connections to the device,
such as accesses to the communication ports and channels (e.g., USB, PCIe~\cite{usb1, usb2, pcie1, pcie2}), or direct tempering the motherboard (e.g., via soldering)~\cite{mb1, mb2}.

\item \textit{Physical side-channel attacks} focus on leaking secrets based on physical behaviors of the components during security sensitive workloads, including power analysis~\cite{power1, power2}, electromagnetic analysis~\cite{electro1, electro2}, and so on.
Launching such attacks requires an attacker to access the target device physically or through malicious apps, which is highly feasible under edge computing, where the attack can be conducted in an isolated and stable environment.
\end{enumerate}

\PP{Physical protection boundary at edge}
The cloud servers' physical access control models do not apply to edge devices due to the high volume and diversity.
Not to mention the procedure of access control needs to be audited by the cloud server
from remote. 
Hence, an adequate physical access control model tailored for edge computing is under urgent call.

\subsection{Insecure Hardware Components}
\label{ss:hw}
\begin{enumerate} 
\item A common source of insecure hardware components stems from
the broken chain of trust of the hardware components, such as due to an untrusted supply chain~\cite{supply-chain1, supply-chain2} or lack of unforgeable Hardware Security Modules (TPMs)~\cite{tpm1, tpm2, tpm3} and Hardware Security Modules (HSMs)~\cite{hsm, hsm2}.
Such vulnerabilities might hide in the original hardware package and are difficult to mitigate thoroughly without replacing the whole flawed component.

\item Other hardware attacks exploit the inherent design of the components that are critical to its correct functionalities and thus are even more challenging to mitigate holistically.
    \begin{itemize} 
    \item \textit{Energy attack}~\cite{energy} aims to render the device inoperable by draining the equipped batteries through excessive legitimate operations.
    Such an attack can be launched across different layers of a device stack, including hardware resources (e.g., GPS, sensors, and related operations), software resources (e.g., system calls, API, memory allocation, locking), network operations (e.g., data transfer, handshaking protocols, the bandwidth, and antenna).
    
    \item \textit{Rowhammer attack}~\cite{rowhammer1,rowhammer2}, 
    tries to trigger random bit flips in the RAM/DRAM via the electronic interference of neighboring memory cells to tamper with security-sensitive states (e.g., access control bit, root bit, etc.).
    
    \item \textit{Covert channels}~\cite{covert} have drawn increasing attention as they exploit the unintended communication channels stemming from the normal operations of the shared components such as DRAM and Last Level Caches (LLC). 
    They can break the data protection mechanisms enforced by the system to restrict unintended communications and provide the malicious applications a stealthy way to transfer (security-sensitive) data between each other.
    
    \item \textit{Mircoarchitectural side-channel attacks} exploit secret related micro-architectural events, including those inside CPU caches (e.g., Prime+Probe~\cite{primeprobe}, Flush+Reload~\cite{flushreload}), and Translation Lookaside Buffers~\cite{tlb}, branch predictors with speculative executions (e.g., Spectre and Meltdown~\cite{spectre, meltdown}).  
    \end{itemize}
\end{enumerate}

\PP{Edge hardware} Attackers under edge computing environments
are even more advantageous in hardware attacks.
Such attackers directly possess the devices and can perform physical attacks within an isolated and stable environment tailored for their needs, making traditional hardware attacks more feasible. 

\subsection{Insecure Firmware}
\label{ss:firmware}
A computing platform delegates the complexity of hardware initialization tasks to the earlier boot stage using firmware in order to simplify the operation system code.
Under edge computing, hardware and firmware's diverse and proprietary nature proliferates the fear of effective widespread exploitation.

\begin{enumerate}
\item \textit{Firmware modification attacks}~\cite{firmware1, firmware2}
aim to inject malicious logic into the target device firmware.
Usually, such attacks are achieved via firmware update features instead of directly exploiting flaws in the software.
Firmware update is a common feature in most modern systems, while not ubiquitously protected by sufficient security measurements.
Lack of security checks (e.g., signature verification) prior to firmware updates directly facilitates successful firmware modification attacks.
A secondary payload following the successful exploitation of the device
via traditional attack vectors, such as memory modification attacks~\cite{memmod1, memmod2}, can be used to bypass checks if they exist.
The malicious code running at the firmware level could be used to compromise any components that are loaded later in the boot process,
such as the boot loader and the OS (or hypervisor).

\item \textit{Hard-coded and weak passwords} in firmware~\cite{firmware-analysis} is another major concern, especially in devices that embed default passwords while lacking administrative management.
%
%
Although the use of hard-coded or weak passwords can save the maintenance overhead, it leaves the devices vulnerable to naive password-based attacks such as dictionary attacks. Such attacks are extremely easy to conduct using existing tools (e.g., John the Ripper~\cite{jtr}, HashCat~\cite{hashcat}) without domain-specific knowledge.


\end{enumerate}

\PP{Firmware in edge hardware} Besides the threats to the integrity of the firmware, edge devices with legitimate but outdated firmware are also in the attackers' interest~\cite{stale1, legacy1, rollback1}.
On the one hand, because of the different form factors of devices, the diversity of environment for devices in the field, and the specific management requirements, legacy devices cannot be updated in time
and therefore expose known vulnerabilities to attackers. 
%
On the other hand, as the attacker under edge computing
has more authority over the operation environment
(e.g., by directly possessing the device or by less strict physical access control)
, the firmware updating process could be spoofed (e.g., via MITM) to preserve a stale version.
As a result, edge computing may magnify the impact of legacy firmware attacks
due to the attacker's overseeing of the firmware management process and a much larger amount of stale devices.

\subsection{Insecure System Software}
\label{ss:system}
When the OS and the system software are trusted, i.e., the requests from users are faithfully fulfilled, traditional software vulnerabilities can still exist.

\begin{enumerate}
\item \textit{Memory corruptions}~\cite{mem1, mem2, mem3}
(e.g., memory overflow, improper boundary check, lack of sanitization,
double-free, use after free) can lead to control flow hijacking (e.g., ROP~\cite{rop}), and result in violations of the integrity and confidentiality of system data, or even grant complete access to the system (e.g., root-access).

\item \textit{Race conditions} in the system can lead to TOCTOU attacks~\cite{toctou} and lead to similar consequences.

\item Even with a formally verified OS~\cite{formally1, formally2}
that are free of such vulnerabilities, \textit{flaws in the device drivers and third-party libraries} can still completely break the established security guarantees~\cite{lib1, lib2, driver1, driver2}.

\item Besides traditional software vulnerabilities, \textit{improper access control implementation} allows the attacker to gain access to sensitive data without launching an end-to-end exploit~\cite{access1, access2} and generate data flows that are otherwise disallowed.

\item \textit{The system software cannot always be trusted}, as the devices are directly possessed and provisioned by untrusted administrative parties.
When facing privileged attackers such as malicious OS and hypervisors, Trusted Execution Environments (TEEs) always come into play to isolate the security-sensitive content~\cite{sgx, sev, trustzone, keystone}.
However, privileged attackers can still use \textit{side-channel attacks} to study the program runtime behaviors, such as the through control flow~\cite{controlled1, controlled2} or the memory access patterns~\cite{cache1, cache2}, and leak the secret, as the OS is still responsible for management tasks such as handling page faults.
\end{enumerate}

\PP{SELinux and Security Monitors}
Edge devices and servers that enable defensive features SELinux~\cite{selinux} and security monitors~\cite{monitor} may temporarily turn off these features for power and performance considerations, leaving chances for attacks.
Not to mention that the features are not free of vulnerabilities~\cite{selinux-vuln}.
An untrusted OS can maliciously manage the edge devices,
and temper with the overall infrastructure of an edge computing entity.
\section{Network and Communication} 
\label{sec:netsec}
Looking into the network-level security concerns of edge computing, one must deal with various threats and vulnerabilities pertaining to the communication among network nodes. We first discuss network security issues that emerge from network-related vulnerabilities found in devices connected in the edge network (\ref{sec:dev_vuln}) and edge services (\ref{sec:edge_vuln}). We then discuss issues with the protocol (\ref{sec:prot_vuln}) when establishing trust (\ref{sec:trust}) and proving that an edge node is delegated to perform computation tasks (\ref{sec:cmpute_auth}). Finally, we discuss side channel (\ref{sec:side_chan}) and Denial of Service attacks (\ref{sec:dos}), and outline research challenges (\ref{sec:netsec_research}).

\subsection{Device Vulnerabilities}
\label{sec:dev_vuln}

The physical layout of the edge network introduces new threats. One such threat is the physical access to the machine itself. Edge nodes are physically located close to where data are generated or need to be processed. Thus, it is beyond the vendor's or service provider's physical control. A threat actor may leverage physical access to the device to perform various activities. For example, they could cause Denial of Service (DoS) by damaging or unplugging the device or attempting to penetrate the system by connecting to an open port, or replacing the device with a rogue one. Wiretapping is also possible, allowing packet sniffing or even injection. 

In real-world deployments, devices are highly heterogeneous and may be developed by different vendors. This is an undebatable fact, especially in the context of IoT. Such devices are often found running vulnerable code (or firmware) that has not received proper security assessments. Some devices may have common vulnerabilities, such as out-of-bounds-write, which adversaries can leverage. For instance, Philips Hue Bridge (model 2.X) contains a heap-based buffer overflow which allows remote code execution. Often, the device may utilize buggy libraries even though they could be well-known. OpenSSL library is a widely used cryptographic library that implements secure communication protocols such as SSL/TLS. According to CVE-2014-0160~\cite{heartbleed}, the heartbleed bug in OpenSSL's heartbeat implementation leads to memory leakage from the server to the client and from the client to the server. Such leakage could reveal secret keys or other protected material.

Network nodes such as Network Address Translators (NAT) carry vulnerabilities of their own. A NAT ``hides'' the IP addresses of internal network devices from the outside network. All internal devices share one public IP address. Thus a threat actor from the outside cannot know which devices exist in the internal network. However, the slipstreaming vulnerability found in NAT~\cite{natslip} actually allows outsiders to learn which devices are in the internal network and probe them. The adversary only needs to convince one of the devices in the internal network to connect to a malicious domain. 

\subsection{Vulnerabilities in the Edge Services}
\label{sec:edge_vuln}

An adversary may be able to inject a malicious payload using several underlying techniques such as SQL injection\cite{halfond2006classification}, XSS~\cite{wassermann2008static}, and CSRF~\cite{barth2008robust}. The main root cause of all those techniques is the lack of proper input sanitization, where the edge server needs to check the validity of the input request. Often, servers merely check only for correct syntax, which is trivial. However, they do not reason about the contents of the request, e.g., code is found where there should be data or vice versa. For instance, a vulnerability was found in the Cisco Fog Director that allowed an unauthenticated attacker to conduct an XSS attack against a user of the web interface of the affected software~\cite{ciscofog}.

\subsection{Protocol Vulnerabilities}
\label{sec:prot_vuln}

The Edge network usually uses lightweight communication protocols such as LTE~\cite{sesia2011lte}, Wi-Fi, Bluetooth, MQTT~\cite{standard2019mqtt}, CoAP~\cite{shelby2014constrained}, AMQP~\cite{vinoski2006advanced}, LoRa~\cite{loraspec}, and Zigbee~\cite{zigbee}. Cloud computing employs heavyweight protocols such as TLS~\cite{dierks2008transport}, HTTPS~\cite{rescorla2000http}, and FTP~\cite{postel1985rfc0959}. The protocol itself becomes part of the attack surface. Soft spots in the communication protocols may provide an attacker the grounds to launch an attack. 

The Zigbee (before version 3.0) required communicating devices to share a pre-master secret key. This key was installed on devices by vendors. However, considering the millions of IoT devices, an adversary can inevitably figure out the pre-master secret~\cite{wang2020analyzing}. Another possibility is that an adversary may force the endpoints to downgrade the security of the communication~\cite{moller2014poodle}. Such an attack is possible if the endpoints have no way of verifying that the security properties agreed are the ones that were truly intended. M{\"o}ller et al. , demonstrated the POODLE attack (Padding Oracle On Downgraded Legacy Encryption) on SSL 3.0, which allows stealing secret material, such as HTTP cookies. HTTP is used in the edge-cloud communication.  

Message Queuing Telemetry Transport (MQTT) is an application-layer communication protocol widely used for IoT to edge communication. While MQTT protocol supports encrypted communication, it is optional. This configuration could result in critical privacy violations while allowing a man-in-the-middle to inject messages. For instance, data generated from wearable devices could include highly sensitive medical data, personal information, and even people's movements.

Like MQTT, CoAP is another application layer for edge devices and applications that works on top of UDP. It has been reported that CoAP is susceptible to attacks such as address spoofing and amplification attacks~\cite{coap_online, arvind2019overview, coap_ietf}. In CoAP the response packet is much larger than the request packet, and thus an attacker can use small UPD requests to generate large-size responses from CoAP nodes, thus causing denial-of-service to the victim devices (see section 11.3 in RFC 7252~\cite{shelby2014constrained}).

\subsection{Establishing Trust}
\label{sec:trust}

\textit{Authentication} poses another challenge. Low-end devices may be authenticated to the edge servers using weak credentials. Thus, an adversary may even be able to perform a dictionary attack to break into the system~\cite{dictionary_attack}. However, vulnerabilities may be introduced due to the usage of weak cryptographic authentication protocols (e.g., WEP~\cite{stubblefield2002using}) or unpatched versions of them (e.g., WPA2-PSK dictionary attack~\cite{nakhila2015parallel}). Authentication code-level implementation poses additional threats to entity authentication. Code that has not been tested or peer-reviewed may implement authentication in a wrong way, granting access to unauthorized entities. A good example of that is Apple's ``goto fail'' bug, which allowed a MITM adversary to compromise the end-to-end secure TLS connection. The bug was essentially bypassing the certificate verification of the server, thus compromising the authenticity of the connection~\cite{apple_ssl_bug}. 

\subsection{Compute Authorization}
\label{sec:cmpute_auth}

One crucial contribution of edge computing is the delegation of complex computing tasks to the network's edge. While authentication aims to solve the problem of verifying identities, authorization deals with the problem of verifying whether a particular node is authorized to perform a particular computing task. A similar problem is encountered in Content Delivery Networks (CDNs), where media stream providers (e.g., Netflix) authorize servers to deliver content on its behalf in several regions. The same physical spread applies in edge computing as well. Similar to how end-users need to verify the content received from a CDN server, an edge device needs to be able to prove that it is authorized to perform such computation from the core network (e.g., corporate network, service provider).

\subsection{Side-channel Attacks}
\label{sec:side_chan}

When the attacker can gain knowledge about the occurring communication and the endpoints communicating, they may be able to use this information to attack the infrastructure itself. Information leaked through side channels often reveal information about the secret keys. Ronen et al.~\cite{ronen2017iot} demonstrated a novel side-channel attack on Philips Hue smart lamps, which revealed the initialization vector and the secret key that the lamps use to authenticate and encrypt the firmware. Thus, information gained from the side channel could lead to catastrophic results. In addition, side channels raise privacy issues, especially in environments where privacy preservation is of high importance (e.g., smart home).

\subsection{Denial of Service Attacks}
\label{sec:dos}

Edge computing is a complex infrastructure that includes the interconnection of edge devices, edge servers, and the cloud. As discussed earlier, edge device vulnerabilities or vulnerabilities in the protocol itself may allow an attacker to cause disruptions. The edge network is more susceptible to distributed denial of service (DDoS) attacks since it contains computationally less powerful resources than the cloud. For the same reason, often, services deployed at the edge are error-prone in their security settings~\cite{xiao2019edge}. For instance, if an attacker can take control of a cluster of edge devices, they essentially create a botnet. A very famous attack is the Mirai botnet~\cite{kolias2017ddos} where attackers were able to compromise IoT devices and use them to cause a DDoS to the edge server network providers such as Krebs and OVH~\cite{antonakakis2017understanding}. Then, the botnet floods the core network (i.e., the edge servers) with enormous requests. Such an attack causes a DoS due to resource exhaustion. Techniques to cause a denial of service include message flooding at the internet layer (e.g., ICMP), the transport layer (e.g., TCP, UDP), or even the application layer (e.g., HTTP).

While DDoS attacks are more practical in the edge network, cloud DDoS is still feasible from the attacker's perspective. Successful DDoS attacks on the cloud can cause significant disruptions in the edge network. CVEdetails report the top 50 vulnerable products~\cite{cvetop50}, many of which run on cloud servers (e.g., Windows 10 and Linux Debian).

\subsection{Research Challenges}
\label{sec:netsec_research}

From the network security point of view, many challenges need to be addressed in the future research. 
\begin{enumerate}   
    \item \textit{Network security configurations and management:} Edge computing is designed to support a range of devices spanning different vendors with different security capabilities. Different vendors provide different application programming interfaces to configure the device. Administrators must make significant efforts to adequately configure the devices, making it a costly and error-prone task.
 
    \item \textit{Adoption of Zero Trust Architecture (ZTA):} As argued earlier in the section, access control is critical when managing large distributed infrastructures like edge computing. Frameworks are needed to support fine-grained access control policy specifications; in other words, which communications should be allowed and why. Defining such access control policies needs to be done in a way that keeps errors minimal. For example, an administrator may specify at a high level which entities need to communicate and have an automated process to translate this to network-level details (e.g., using firewall rules, etc.). Often, organizations adopt threat models with weak adversarial assumptions. For instance, they may assume that the adversary may not be able to compromise end devices. Therefore, access control is very loosely defined within the network while enforcing more access control on network perimeters (i.e., firewalls). However, this becomes a major issue once the presumably trusted devices misbehave. For example, a compromised IoT device may start probing resources on the network or try to spread the infection further in the network. Such an attack is feasible since access control is non-existent or very loosely defined~\cite{ronen2017iot}. In other words, one needs to precisely define who can access what on the network and for what reason~\cite{katsis2021can, katsis2022neutron}. 

    \item \textit{Secure access service edge (SASE):} SASE extends the ZTA by granting access to network resources based on the entity's identity (e.g., IoT, device, user application), while it also requires real-time context and continuous assessment of risk and trust throughout the sessions. Real-time contexts can vary across edge computing deployments and must be standardized for compatibility across SASE services. Assessment of risk and trust must be a continuous monitoring process as long as an entity lives in the network. However, as we discuss in the paper, risk and trust must account for threats and vulnerabilities emerging from different architectural stacks (e.g., network, hardware, system). 

\end{enumerate}
\section{Cloud Stack}


The cloud composes the central processing and maintenance of the data collected in the edge network. The edge network transmits data to the cloud for further processing or permanent storage, among other purposes. The cloud is a distributed system composed of several interconnected machines similar to the edge network. However, cloud resources are shared among many parties. In public clouds, those parties can be different users or organizations. Therefore, security issues emerging in the cloud impact edge network as well. Despite its architectural incentives and benefits, the cloud brings its security issues to the edge network.


\subsection{Cloud Vulnerabilities}
\label{sec:cloud_vuln}

\begin{enumerate}

\item \PPi{Misconfigurations:} Data can be exposed simply by misconfiguration of the cloud machine instances (see Figure~\ref{fig:edge-arch}). For example, extremely sensitive user data were leaked by a cloud leader due to a misconfiguration of Amazon Web Services S3~\cite{accenture, verizon}. The misconfiguration allowed public access to the cloud server hosting the data while also having no access control enforcement. 

\item \PPi{Insecure APIs:} Cloud services expose their functionality through Application Programming Interface paradigms (e.g., REST). Gartner predicts that by 2022, application programming interface (API) attacks will become the most-frequent attack vector, causing data breaches for enterprise web applications~\cite{checkpoint}. The most critical issues in API security are insufficient access control, injection, and Excessive Data Exposure, among others~\cite{owasptopapi}. A vulnerability in Microsoft Exchange Server allowed attackers to send unauthenticated HTTP requests to any Exchange server. Broken user authentication and security misconfigurations allowed the adversaries to leverage the back-end API of the server to escalate privileges and maintain persistence. A similar scenario could be encountered in an edge computing environment. Services deployed at the edge expose API functionality to be used by devices.  

\item \PPi{Virtualization issues:} Virtualization operations can be distinguished into two components: Virtual Machine Manager or Hypervisor (VMM) and the Virtual Machine instances.

VMM is the crown jewel since its compromise allows the adversary to attack many virtual systems at once. In 2021, a ransomware attack occurred on the VMware ESXi hypervisor. According to Sophos firm, defensive mistakes and unnecessary functionality allowed the attackers to deploy crypto-locking malware and disrupt the operations of all the tenant VMs.

VM integrity must be verified at all times in the cloud. VM could be running mission-critical applications for the edge network; thus, cloud providers must ensure that there are no malicious or compromised VM images. In addition, vulnerabilities associated with virtualization environments may allow attackers to obtain administrator privilege in the host system due to mishandling of privileges (CVE-2019-5736~\cite{CVE-2019-5736}).

\item \PPi{Cloud Service Provider transparency:} Often, there is no clear understanding of how the cloud operations are performed, and thus, it is not feasible to assess the security posture of the edge network thoroughly.  For example, it is unclear what protection measures the cloud provider takes for data confidentiality. How are the data stored, or when do they get decrypted for processing? Are there any process isolation mechanisms in place? Those are just a few of the security concerns. Of course, similar considerations apply to data integrity and availability. For example, when does the signing process takes place and where? What is the mechanism employed for data availability? Another interesting aspect is if third parties are involved in the cloud operations, for instance, for data backups. Transparency is the key factor for clients and network administrators to realize risks and see how things can go wrong.

\item \PPi{Cloud at the Edge:} Cloud at the Edge (aka Edge Cloud) offers cloud computing resources to the edge of the network or where the traffic is. Various frameworks have been developed, offering cloud services at the edge, such as KubeEdge~\cite{kubeedge}, EdgeX Foundry~\cite{edgeX}, and OpenEdge~\cite{openedge}. Such frameworks are part of the cloud stack and bring their flaws into the edge infrastructure. A recent vulnerability was disclosed where specific binaries within the OpenEdge application were susceptible to privilege escalation (CVE-2022-29849). A local attacker could elevate their privileges and compromise the affected system. Similarly, due to improper access control and weak password requirements in EdgeX, an attacker could make authenticated API calls to EdgeX microservices from an untrusted network.

\end{enumerate}

\subsection{WebAssembly (Wasm)}

Wasm~\cite{wasm} was proposed by the World Wide Web Consortium (W3C) as a platform-independent compilation target for various high-level languages (e.g., C, C++, Rust).
Originally, Wasm addressed the problem of safe, fast, portable low-level code on the Web.
As an abstraction over modern hardware, Wasm is language-, hardware-, and platform-independent, with use cases beyond the Web.
Wasm adopts a linear memory with a configurable size for all memory accesses other than local and global variables.
The liner memory region is disjoint from other memory regions (e.g., code space, execution stack), containing the impact of vulnerabilities within the data of the program's own memory.
\begin{enumerate}
\item Beyond all the benefits brought by Wasm, research has shown that traditional vulnerabilities re-instantiate in Wasm~\cite{wasm-sec1, wasm-sec2}.

\item
Even worse, Wasm enables unique attacks, such as overwriting constant data or manipulating the head using a stack overflow.
Attack primitives found in Wasm include but are not limited to those that allow an attacker: 1) to write arbitrary memory, 2) to overwrite sensitive data, and 3) to trigger unexpected behavior by tampering with control-flow integrity (CFI) or the host environment.

\item \PPi{Threats to Wasm at edge}
Researchers have explored the application of Wasm under the setting of edge computing~\cite{wasm-edge1, wasm-edge2, wasm-edge3}.
As Wasm is designed as platform-independent, exposed vulnerabilities will remain in edge devices that adopt Wasm, with potentially higher impacts by propagating the outcome across the massive and heterogeneous edge computing network.
\end{enumerate}
\section{Machine Learning Workload}
\label{sec:ml-workflow}


With the increasing amount of data generated by a large number of devices in the edge network, the opportunities, challenges, and applications of edge machine learning (edge-ML) have also increased. 
Edge-ML has been widely used for various edge computing tasks, e.g., computer vision for traffic surveillance~\cite{bozcan2021context}, decision making for autonomous driving~\cite{wang2019pseudo}, and speech recognition for personal assistance~\cite{kumar2018skill}.
Considering the huge number of devices in the edge network, training edge models from scratch will require tons of computational resources. 
Therefore, transfer learning~\cite{torrey2010transfer} along with other techniques~\cite{merenda2020edge} is introduced to reduce the development cost of edge-ML systems~\cite{hou2018proactive,daga2019cartel}.
In this section, we discuss the edge-ML system security from several aspects, including life cycle (\S\ref{sec:ml-lifecycle}), threats (\S\ref{sec:ml-threats}), vulnerabilities and exploits (\S\ref{sec:ml-vuln-exploit}), and challenges (\S\ref{sec:ml-challenges}).

\begin{figure}
\centering
\includegraphics[width=\linewidth]{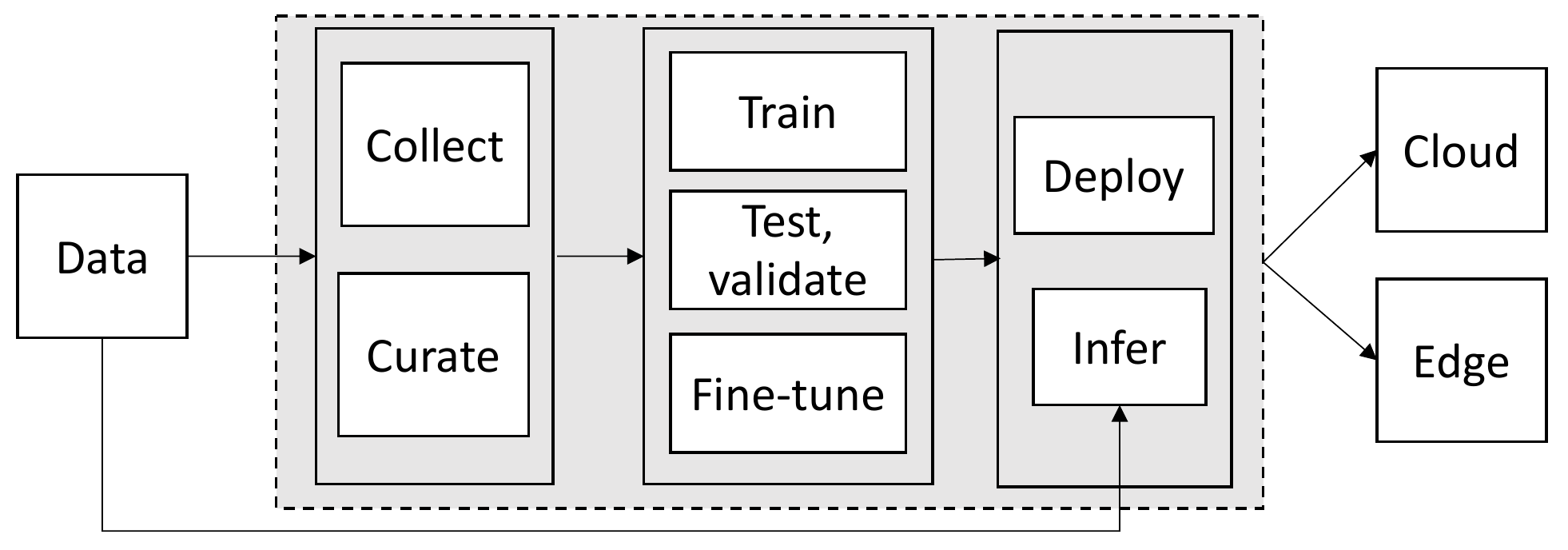}
\caption{The Life Cycle of Edge Machine Learning Systems}
\vspace{-0.2in}
\label{fig:edge-ml-lifecycle}
\end{figure}

\subsection{Edge-ML Life Cycle}
\label{sec:ml-lifecycle}
\autoref{fig:edge-ml-lifecycle} shows the life cycle of edge-ML systems, which includes stages of data collection and curation, model training, testing and validation, deployment, and inference. 
The key feature of edge-ML is allowing model training across decentralized edge devices or servers that hold local data samples without exchanging them, which is also referred to as \textit{federated learning}~\cite{li2020federated,wang2019adaptive}.
The diverse edge computing tasks atop different edge environments and the new learning paradigms bring about a set of domain-specific security problems and challenges of edge-ML.
Building trustworthy edge-ML systems requires securing the whole life cycle of edge-ML systems in terms of the data, model, and infrastructure.


\subsection{Edge-ML Threats}
\label{sec:ml-threats}

In edge-ML, the availability and visibility of local data depend on the two different training patterns: (1) local data are collected and uploaded to servers while the models are trained in a centralized way and (2) local data is only available to the local workers (edge devices) in federated learning scenarios.
For the first training pattern, all the known threats~\cite{liu2018survey} for centralized machine learning systems can be applied.
For the second training pattern, there are new threats due to the nature of distributed learning. Distributed learning makes it difficult to audit the quality of local data and the training behaviors of local workers.
Malicious local workers can manipulate the final models by modifying the training data in silence.
In addition, attacks can also be performed on the data collection, transmission, and processing phases by attacking benign workers~\cite{abdulrahman2020survey}.
To make it clearer, we define the threat model with two types of attacks: insider and outsider attacks~\cite{walton2006balancing,rodriguez2022survey}. 
When attacks are carried out by any participant (local workers or servers) in the decentralized learning system, they are defined as \textit{insider attacks}.
The \textit{outside attacks} are mainly carried out by sniffing/inferring information about the data or models.   \looseness=-1

\subsection{Edge-ML Vulnerabilities}
\label{sec:ml-vuln-exploit}

The new threats to edge-ML systems introduce new vulnerabilities by which adversaries can further develop exploits to attack systems. Based on the life cycle of the edge-ML system, we consider vulnerabilities and exploits in different steps of this system.
In this section, we skip the discussion of attacks targeting data collection and curation (we present them in \S\ref{sec:ml-data}) to differentiate between direct and indirect attacks.    \looseness=-1

\PP{Training-time attacks} For training-time attacks, the adversary can manipulate the training data to perform poisoning attacks, injection attacks, adversarial perturbation attacks, and byzantine attacks.    \looseness=-1

\begin{enumerate}
    \item \textit{Poisoning attacks}~\cite{gu2017badnets,shafahi2018poison,jagielski2018manipulating} aim to poison the training datasets by adding a small number of poisoned samples (e.g., 5\% of all training samples~\cite{gu2017badnets}). The data poisoning process can be achieved by providing wrong samples, injecting perturbations, or flipping labels. For example, the attackers can add fake data via SSD attacks~\cite{baek2020ssd}, compromise the edge software systems and devices~\cite{xiao2019edge}, and IoT spoofing~\cite{hasan2019protecting}. The victim edge-ML systems will learn the erroneous features and update the model weights incorrectly.
    \item \textit{Injection Attacks}~\cite{tang2020embarrassingly,li2020backdoor} are prosecuted by inserting trojans and bias to only the target data sample, which is often called target attacks. The triggers can be physical objects (e.g., stickers on the stop sign~\cite{eykholt2018robust}) or the digital triggers (e.g., watermarks on images~\cite{shafieinejad2021robustness}), which can fool the edge-ML systems (e.g., autonomous vehicles and face recognition cameras) to predict inputs with triggers as the target wrong labels.
    Moreover, attackers can also add bias, resulting in fairness problems~\cite{mehrabi2021survey} to victims. For example, they can manipulate the edge sensor data to force the classifiers to learn unbalanced labels (e.g., gender~\cite{leavy2018gender}).
    \item \textit{Adversarial example attacks}~\cite{yuan2019adversarial,xiao2018generating} are performed by adding perturbations into training samples, which are carefully generated and indistinguishable to the human eyes. The attackers can actively generate dynamic and optimized perturbations to mislead machine learning systems. For example, the adversary can attack the edge network intrusion detection systems by adaptively mutating the network features (e.g., device IP address and port number) by DDoS, reconnaissance, and information theft attacks~\cite{koroniotis2019towards}. 
    \item \textit{Byzantine attacks}~\cite{fang2020local,so2020byzantine} are carried out when the cloud or edge servers select malicious clients in the federated learning stage. Attackers can contaminate the local edge device data via poisoning attacks~\cite{tolpegin2020data} and use it to train the local models. Then they upload the well-trained or directly tampered malicious model weight updates to servers~\cite{baruch2019little,shejwalkar2021manipulating}, which can damage the final model hosted in the servers or leave a backdoor~\cite{bagdasaryan2020backdoor}.
    Except for malicious model weights, the attackers can also directly upload the malicious local data (crafted by data manipulation~\cite{hu2021challenges}) to the servers. When data security checks fail to identify these vulnerabilities in servers, attacks will succeed~\cite{li2021byzantine}.
\end{enumerate}

\PP{Inference-time attacks} Inference-time attacks can be performed in the stages of offline validation/testing and online inference. We consider four categories of inference-time attacks, including exploratory attacks, membership inference attacks, evasion attacks, and spoofing attacks. 
\begin{enumerate}
    \item \textit{Exploratory attacks}~\cite{shi2018vulnerability,tramer2016stealing,papernot2017practical,wu2016methodology} aim to explore the architectures of the machine learning system and build shadow models (or edge-ML systems). 
    The adversaries (e.g., malicious edge devices) have black-box access to the victim model (e.g., the final model) through MLaaS APIs~\cite{ribeiro2015mlaas}, where edge or cloud servers provide machine learning service APIs to edge devices. Attackers first query victim models with synthesized samples~\cite{papernot2017practical,wu2016methodology} to get labels and then train a functional equivalent~\cite{tramer2016stealing} shadow models based on the labeled data. \looseness=-1
    \item \textit{Membership inference attacks}~\cite{nasr2018machine,leino2020stolen,song2019privacy} are performed to determine if the given data samples are members of the victim models' training data.
    In edge-ML, overfitting final models force them to memorize the training samples, which can increase the success rate of this attack~\cite{song2019privacy}.
    \item \textit{Evasion attacks}~\cite{shi2017evasion,papernot2016limitations,canim2019uncheatable} are to manipulate inference results by carefully crafting test samples. 
    The adversary can be either insider attackers or outsider attackers that compromise the edge devices~\cite{taneja2013analytics} or invade the edge communication network~\cite{lee2016analysis} to generate the samples that have the close probabilistic distribution as the original training data to fool victim models, resulting in prediction errors. 
    \item \textit{Spoofing attacks}~\cite{shi2019generative,shi2020generative,luo2020adversarial} are performed by directly generating adversarial test samples from scratch rather than crafting existing samples. The adversary (e.g., malicious edge devices) can spoof edge device data by GAN networks~\cite{goodfellow2014generative}, which consist of two components: generators and discriminators. Generators create spoofing signals, and discriminators detect whether samples are spoofed or not, where they minimize their own low by playing minimax games~\cite{koolen2014efficient}.
\end{enumerate}

\PP{Deployment attacks} After the training, validation, and testing stages, edge-ML models will be deployed in edge networks, where they can be placed in edge/cloud servers or edge devices. We consider two categories of deployment attacks:  model stealing attacks and supply chain attacks.

\begin{enumerate}
    \item \textit{Model stealing attacks}~\cite{sun2021mind} can occur in both cloud/edge server and edge device sides. For the cloud servers, the edge-ML models are usually invisible to attackers except for the scenarios where attackers can invade the servers. However, they can still indirectly steal the model by the above-mentioned inference-time attacks (e.g., exploratory attacks). When the edge-ML models are deployed in the edge servers and devices, it is much easier to get the models due to the lack of model protection techniques. Sun et al.~\cite{sun2021mind} find that 81\% of the edge-ML models can be simply extracted from edge applications and directly used. Moreover, most of the edge-ML models are built atop popular and publicly available machine learning frameworks (e.g., Tensorflow Lite~\cite{david2010tensorflow})~\cite{almeida2021smart,xu2019first}, which means that attackers can easily reuse the stolen models. Although many countermeasures, e.g., model deployment with trusted execution environments~\cite{mo2021ppfl,hashemi2021darknight}, have been proposed, the limited resources of edge devices make them far from practical.
    \item \textit{Supply chain attacks}~\cite{he2021model,omitola2018towards,yang2015protecting} are performed on the edge-ML models distributed in the supply chain. As deep neural networks become deeper and deeper, edge-ML models are often built on existing pre-trained models (e.g., mobileNet~\cite{howard2017mobilenets}) that are specifically designed for edge computing environments via transfer learning. In this case, edge-ML models need to be distributed on numerous devices for both federated learning and online inference tasks. However, the supply chain of pre-trained model acquisition~\cite{he2021model} and model distribution~\cite{omitola2018towards,yang2015protecting} will be the target of adversaries. For example, the attackers can stealthily modify the weights of models by man-in-the-middle attacks on the edge network. 
\end{enumerate}

In addition to attacks targeting the edge-ML life cycle, there are also threats and vulnerabilities in edge-ML frameworks. 
Unlike machine learning frameworks for cloud servers, which have much larger computational resources than edge devices, the edge-ML frameworks are carefully designed to overcome resource limitation issues. 
The edge-ML frameworks have gained more and more attention recently, and many vendors have designed their own frameworks for the edge environment, such as TensorFlow Lite~\cite{david2010tensorflow}, Apple Core ML~\cite{coreml} and Pytorch mobile~\cite{pytorchmobile}.
Moreover, it has been found that frameworks for general machine learning are also widely used in edge devices, e.g. Caffe2, NCNN, and Parrots~\cite{xu2019first}.
These frameworks used in the edge-ML system can also be vulnerable.
One of the key vulnerabilities of the edge-ML frameworks is the calculation vulnerability~\cite{calculationattack}.
It usually happens when an unexpected data point or value is given to edge-ML frameworks, which involves physical dimensions of edge data, such as size, length, width, and height.
The adversary can develop exploits of this attack to trigger infinite loops or deadlock to break edge-ML systems~\cite{burnim2009looper}.



\subsection{Edge-ML Challenges} 
\label{sec:ml-challenges}

The diverse edge-ML environments bring about a set of challenges in applying machine learning to edge computing. Here we consider the following challenges.

\begin{enumerate}
    \item \textit{Computational challenges}. The edge devices usually have a very limited computational resource budget for edge-ML model training and deployments. For example, the memory on the chip of IoT devices is $10^{3}\times$ and $10^{5}\times$ smaller than mobile devices and cloud servers~\cite{lin2020mcunet}. The limited resources make it hard to deploy the defenses and countermeasures of the ahead-mentioned attacks. Even if the defenses can be applicable for edge computing, the performance of the main tasks will be affected while the defenses are used. For example, as discussed above, although the TEE solution for model protection can be used for some edge devices, TEE enclaves will consume most of the on-chip memory to affect the other tasks on the devices~\cite{mo2020darknetz}.
    \item \textit{Life-cycle protection challenges}. Except for the limitation of computational resources, protecting all stages of the whole life cycle of edge-ML systems itself is very hard because of the large number of unprotected edge devices.
    Taking into account the federated learning paradigm, attackers can easily access local data and models due to the lack of protection mechanisms adopted in edge devices.
    In addition, the communication channel between edge devices and servers can also be vulnerable (as we discussed in \S\ref{sec:netsec}), which also presents challenges to data and model protection. 
    Hierarchical and collaborative protection schemes are required to secure the edge-ML workflow, which is difficult to achieve.
    \item \textit{Ethics challenges}. Although edge computing has become an inevitable part of our society and millions of edge-ML systems and models have been deployed~\cite{alexanumber}, there are several unsolved ethic issues. 
    First, the owner identification issue exists in edge-ML systems. 
    Collecting various user data without user's consent or permission is a critical problem to be addressed.
    Second, there is no clear line to differentiate private and public data, where edge devices can collect both public and private data from end-users. The absence of clearly defined boundaries for data leaves attack surfaces to adversaries, and they can develop exploits to attack users from the edge-ML system.
    Finally, even if there are many countermeasures, e.g., differential privacy, that have been proposed to address some of the problems~\cite{mothukuri2021survey}, edge-ML ethics is still an open question~\cite{loi2019include}.
\end{enumerate}
\section{Cryptography}
Cryptography is a fundamental stone in edge computing security in various aspects ranging from communication and storage to computation and analysis. Due to the difference in available resources in edge devices and servers, the strength of cryptographic protection on each device or server varies. 

In this section, we focus on using cryptographic algorithms in the edge network, define the life cycle of different aspects of a cryptosystem, and analyze potential threats and vulnerabilities we may face in this new computational framework. We then discuss other cryptography topics in edge computing, such as entropy management and post-quantum cryptography.

\subsection{Cryptography Life Cycle}
Cryptography life cycle refers to the different phases of use of cryptography in and across systems, especially for network security. 
The components of the cryptography life cycle include:
key management life cycle,
cryptosystem design and implementation, 
protocol deployment and adoption, 
cryptosystem expiration and revocation, and supply chain.

The key management life cycle refers to the phases of using cryptographic keys, such as key generation, updates, and revocation. 
Cryptosystem management focuses on the implementation, adoption, and retirement of the protocols themselves, whereas supply chains focus on the origin and paths in developing cryptographic libraries and packages.

\subsection{Cryptographic Vulnerabilities}
Based on the cryptography life cycle, we list out potential vulnerabilities and exploits that may be present in each stage of the cryptography life cycle.

\PP{Key management}
Security of cryptographic keys is the most crucial component in a cryptosystem. If an adversary is able to obtain the symmetric key used by an edge device and edge server during communication, all past and future messages encrypted with such key will be under the full control of the adversary. To protect the key, we must be aware of potential exploits during each stage of its life cycle.
\begin{enumerate}
    \item \PPi{Key generation}
        Bit security \cite{mw2018bitsecurity} is a notion of evaluating the security of cryptographic algorithms and is generally related to the size of the key space. Achieving more robust security of a particular cryptosystem typically implies increasing the key length \cite{Lenstra2010KeyLC}, which can slow down the key generation process and limit the number of available keys of a resource-constrained edge device. Using short keys in communication can alleviate the computation burden of edge devices at the cost of security.
        
        Exploits in the key generation stage can also originate from an active outsider adversary. Due to limited resources, an edge device may not be able to properly verify identity of the edge server. During the key agreement process between edge devices and servers (e.g., Diffie-Hellman key exchange algorithm), an adversary can perform a man-in-the-middle attack to obtain the symmetric encryption key \cite{kl2015dhmitm} and tamper with all transmitted messages.
        
    \item \PPi{Key rotation}
        Key rotation \cite{eprs2017keyrotation, googlekeyrotation} is a common cryptographic practice to retire old encryption and signing keys and generate new ones for future communications. It reduces the number of messages each key is linked to, thus preventing adversaries from batch-decrypting and compromising all transmitted messages in the network and providing forward secrecy \cite{yhtz2020keyagreement} guarantee for the protocol. 
        However, frequently generating and deriving new ephemeral keys is resource-consuming and may not apply to many edge devices with limited resources. 
    
    \item \PPi{Key revocation}
        Similar to key rotation, key revocation is also a crucial practice to guarantee forward secrecy and prevent currently unauthorized parties from continuing accessing data within the edge network. It is recorded through key revocation certificate, and a certificate revocation list \cite{certrevocationlist}.
        However, due to a large number of edge devices and servers, and the necessity of regularly refreshing keys, it is time- and space-consuming to keep track of every revoked key for all devices and servers on the network. 
\end{enumerate}

\PP{Cryptosystem management}
\label{pp:cryptosystem}
As the edge network is a new computation system framework, there are frequently new cryptographic schemes \cite{glcxz2018edge, xzpzy2019partially} being proposed to help alleviate vulnerabilities in network security, data protection, and privacy. However, translating from provable theoretical security to software security is shown to be a not easy task.

\begin{enumerate}
    \item \PPi{Design and implementation}
        Studies by Lazar et al. \cite{lcwz2014software} on 269 cryptographic vulnerabilities reported in the CVE database have shown that the vast majority of cryptographic vulnerabilities come from not bugs in cryptographic libraries but misuse of such libraries and packages during protocol and application development. To securely implement new cryptosystems, one need both profound cryptographic knowledge and extensive experience in cryptographic software development, as otherwise it can introduce many unnecessary risks. 

        

    \item \PPi{Deployment and adoption}
        When a new protocol is standardized and ready for public deployment after a careful creation process, we now face the problem of a slow adoption process regarding the new protocol. For example, TLS 1.3 \cite{rfc8446} was introduced in 2018 with significant improvement in performance and security guarantees over the previous version, yet its adoption rate has just reached 63\% by late 2021 \cite{tls1.3adpotion}. The slow deployment process can be caused by public unawareness or the lack of compatibility with old devices and systems. During communication between an edge device and edge server, if an edge device does not support TLS 1.3, the server will have no choice but to downgrade to using less secure cryptographic algorithms and keys, leaving the channel more vulnerable to malicious attacks.

    \item \PPi{Expiration and revocation}
        With the discovery of new attacks and deployment of new cipher suites, certain cryptographic algorithms will be removed from common usage to provide stronger security requirements. For example, TLS 1.3 removed MD5, RSA, and weak elliptic curves from its cipher suite pool \cite{tls1.3difference}. Due to the lack of updates or computational resources, old edge devices and servers may be incompatible to run secure new protocols. Known weak ciphers and hashes such as DES \cite{des, des-cve} and MD5 \cite{md5insecure} may still be used in the edge network, leading to simple and effective attacks \cite{dh1977des, stevens2006md5} against intercepting in-network communications.
\end{enumerate}

\PP{Supply chain} 
\label{sec:supply-chain}
The security of a specific cryptographic library is built upon security assumptions on its dependencies, forming a chain of trust. If one chain link is broken, all packages and libraries that depend on it can be compromised.
Vulnerabilities of the cryptographic supply chain can emerge from many aspects, including cryptographic libraries themselves, their library dependencies, and package developers.

\begin{enumerate}
    \item \PPi{Outdated cryptographic libraries and packages}
        As cryptographic protocols are deployed for decades or replaced over time, developers may not retain frequent monitoring, maintenance, and updates for the packages. Such libraries do not have timely patches and fixes to respond to newly discovered attacks that affect them, making themselves and their dependency successors susceptible to malicious attacks.
    \item \PPi{Insecure dependency sources}
        As mentioned in section \ref{pp:cryptosystem}, the detachment between theoretical cryptographic proofs and real-life cryptosystem development can introduce many vulnerabilities in protocol implementation. Such issues may not be reflected in theoretical analysis of the protocol, since they do not directly originate from the protocol itself. Indeed, physical side-channel attacks such as One\&Done \cite{akdsczp2018oneanddone} analyze the signal activity when performing modular exponentiation to recover RSA secret keys in OpenSSL. 
        The heartbleed bug \cite{heartbleed} is caused by a missing check in the TLS heartbeat extension in OpenSSL, which did not affect other TLS implementations such as GnuTLS \cite{gnutls} and Windows platform implementations \cite{MSheartbleed}. These vulnerabilities reside in the OpenSSL implementation rather than the TLS protocol itself. Such vulnerabilities are not represented in theoretical protocol analyses but have detrimental effects on the actual application of the protocol. 
    \item \PPi{Untrusted developers}
        The third component of the chain of trust falls onto its people -- developers behind the cryptographic libraries and packages. 
        As demonstrated in an empirical study by Blessing et al. \cite{bsw2021empirical}, software developers re-implementing their own cryptographic tools instead of referring to established libraries can produce vulnerabilities at a rate three times as much as with non-cryptographic software. Furthermore, the complexity of cryptographic software has a much larger negative impact on implementation security when compared to non-cryptographic software. In a vast edge computing network, the cryptographic software complexity is substantially higher than other systems. Not using established cryptographic libraries and tools created by a trusted source can impose a much higher chance of introducing additional vulnerabilities into the system \cite{cgk2021cryscanner, cwe1240}. 
\end{enumerate}

\subsection{Entropy Management Vulnerabilities}
Most modern cryptographic algorithms require strong and secure randomness to ensure security against various attacks. Randomness generated from insufficient entropy can lead to serious compromises, such as the prediction of secret keys \cite{Ristenpart2010WhenGR, hdwh2012mining}. Entropy poisoning attacks \cite{Alt2015EntropyPF} can restrict, influence, or give an adversary complete control over the entropy pool used by devices, causing the algorithm outputs to be easily predictable or entirely deterministic in the eyes of the adversary.

\subsection{Quantum Safe Cryptography}
Since data may be stored in edges servers for years or decades from now, one must consider new threats from utilizing quantum computers in the not-so-distant future. 

In the post-quantum age, many popular and traditionally-secure asymmetric algorithms (e.g., RSA \cite{rsa1978}, DSA \cite{dsa2013}, ECDSA, etc.) can be easily broken by a quantum computer running Shor's algorithm \cite{shor1994algorithms}, while the security of symmetric schemes has been drastically decreased by Grover's algorithm \cite{grover1996quantum}. An adversary may be equipped with storage capabilities to store communications records among edge devices and servers, then utilize quantum computers to help decrypt and recover sensitive data after. 
Research has been done \cite{lcg2018securing, fernandez2020quantum, wcxw2022lattice} to investigate methods for building a quantum-resistant algorithm that helps alleviate such issues. However, challenges \cite{fernandez2020quantum} persist in applying quantum-resistant edge devices in many aspects, such as performance, standardization, and physical security. 

\PP{Post-quantum performance concerns}
For a scheme to be considered secure, quantum-secure cryptosystems usually demand a significantly heavier workload on the key generation process \cite{ppb2016shorter} with a much larger key size requirement than in the traditional setting. For example, most symmetric algorithms such as AES \cite{aes} can be considered quantum-safe at the cost of doubling the key length \cite{bernstein2010grover}, but post-quantum asymmetric schemes can have a private key length as large as ~14000 bits \cite{singh2015practical}! Such conditions impose additional computational burdens on resource-constrained edge devices, limiting the scope of their usage and adoption.


\PP{Post-quantum side-channel attacks}
There have been designs and optimizations of lightweight post-quantum algorithms applied to the edge and IoT devices \cite{kmol2019lattice, abtea2019lightweight, ebm2019cryptoprocessors}. Even though such schemes are theoretically secure against quantum attacks, the implementation of quantum-secure algorithms can still be vulnerable to physical side-channel attacks \cite{pskh2018sidechannel, psn2021sidechannel}, as edge devices and servers commonly lack strong physical access control and protection mechanisms.

\subsection{Research Challenges.} 
    The edge computing architecture contains diverse servers and end devices, which raises many challenges when we try to analyze the security of the entire system: 
    \begin{enumerate}
        \item \textit{Cryptosystem configurations.} When numerous end devices and edge servers encompass the cloud, each may support different algorithms, packages, and protocols. It is hard to apply analysis to a general edge computing framework without missing specific details of individual networks.
        \item \textit{Adversarial models.} Different edge-device network configurations can apply different restrictions on the possible models of outsider attacks. One needs to adjust the analysis and mitigation efforts specific to each local network while evaluating the security of the general edge infrastructure as a whole.
    \end{enumerate}
\section{Data Security}


In the edge computing model, a large amount of private data is outsourced to edge servers for computation and storage. 
Information collected from end devices is processed, aggregated, and analyzed in multiple levels of edge servers, then transmitted to the cloud. 
As data travel through the edge server hierarchy, its ownership is transferred from edge devices to many servers. 
This raises questions on how to guarantee the security of a device's data in an untrusted environment without having direct control. 
Furthermore, edge computing aims to provide fast computation and real-time responses to reduce the latency of device-server communication, which presents an additional layer of restriction on applying common data security measures.

We first touch upon different stages of data flow in an edge network and discuss challenges in protecting confidentiality, integrity, and availability during each stage. Then we focus on analyzing data usage issues in machine learning applications on edge.

\subsection{Data Life Cycle}
The data flow within an edge network consists of multiple stages that reflect on different states of the data, including data collection and transmission (data-in-motion), data processing and analysis (data-in-use), and data storage and recovery (data-at-rest). The data shifts among the three states as it travels through the edge network from the cloud. The states interleave with one another, yet each presents unique vulnerabilities in the security of gathered data.

\subsection{General Data Threats and Vulnerabilities}
\label{sec:general-data-vul}

\PP{Data-in-motion}
Data in-motion is the process of data being transmitted to different locations in the network. In edge computing, this includes data collection from various sources, data sharing among edge servers, and data integration from edge servers into the cloud. 
The first step of data flow in an edge network is collecting input data from different sources. Such input can come, for example, from an IoT sensor or other edge devices in the network. Then, the data is shared among edge servers during collaborative computation tasks. In the end, accumulated data is transmitted from edge servers to the cloud.
Ensuring the authenticity and the integrity of data-in-motion warrants the correctness of data as it goes through other stages. 

\begin{enumerate}
    \item \PPi{Weak authentication}
        IoT devices and many edge devices lack sufficient resources to perform advanced authentication algorithms. These operations may be offloaded to the edge server to alleviate the computational burden of edge devices. As a result, data transmitted from devices to edge servers are likely coupled with weak digital signatures \cite{Carminati2018} or message authentication codes \cite{Blanton2009}. In extreme cases, authentication mechanisms may be skipped entirely. This makes incoming messages extremely susceptible to interception and tampering by an active attacker controlling the network. 
    \item \PPi{Fabricated data}
        Even with proper message authentication methods, incoming data may still be sent by a compromised entity. A corrupted edge device or edge server can send fabricated data signed by a valid key to pass verification checks without raising suspicion of other parties on the network.  
\end{enumerate}

\PP{Data-in-use}
Data in-use refers to the phase where data is being processed by the edge server, during which it may be decrypted into plaintext form or remain encrypted for certain operations. At this stage, data confidentiality may be violated by an outside attacker manipulating the server's memory units, or an untrusted server extracting partial information.

\begin{enumerate}
    \item \PPi{Memory-based attacks}
        During the processing stage, stored data is usually decrypted before being used, which provides adversaries with an opening to access decryption keys or plaintext data. In the context of edge computing, since edge servers commonly lack strong physical protection, an attacker with physical access to the edge server can launch memory attacks such as installing RAM scraping malware \cite{ramscraping}, executing untrusted functions \cite{qzagal2020stackvault}, and exploiting side channels \cite{yitbarek2017cold, bm2006cache}.
    \item \PPi{Encryption leakage attacks}
        In cases where the edge server is operating over encrypted data (e.g., via searchable encryption), information about the underlying plaintext can still be exposed to the server via leakage. When performing search, update, and retrieval requests from edge devices, an honest-but-curious edge server can record and analyze memory access patterns to determine the number and frequency of the files accessed. Further, an active server can influence user requests and recover partial plaintext of the encrypted data via leakage abuse \cite{cgpr2015leakage}.
\end{enumerate}

\PP{Data-at-rest}
Usually, data is encrypted while being stored in the edge servers. However, this does not exempt it from potential exploits. 

\begin{enumerate}
    \item \PPi{Data remnants and secure deletion}
        Even though data is encrypted and safely stored on a disk or removed from the server, parts of it may remain in memory from the processing stage if the memory address has not been overwritten. An adversary can target and recover such data remnants via cold boot attacks \cite{gm2013coldboot}.
    \item \PPi{Data backup and recovery}
        Since edge servers can possess sufficient resources to perform relatively intensive computation on collected data, data may not be forwarded to the cloud until the task is done. If stored data is corrupted before being processed and forwarded to the cloud server, it can cause severe data loss and service interruption within the network. This poses an incentive for attackers to target edge servers and inject ransomware \cite{lsct2020selfrecovery, lei2020integrating} in exchange for file decryption keys.
\end{enumerate}

\PP{Data security} Cloud providers may implement weak security practices or have no data protection provisions. Often data are processed or stored in clear text, risking exposure in case of compromise (e.g., the case of Accenture~\cite{accenture, verizon}). 

Data deletion is another critical issue. Research has shown that data persists in memory if not using secure deletion techniques~\cite{yitbarek2017cold}. Also, due to data replications and backups, incomplete data deletion processes might not erase the data in the cloud infrastructure completely.

\subsection{ML Data}
\label{sec:ml-data}

In edge machine learning (edge-ML) systems, the key characteristic is that data provision on edge devices and edge/cloud servers can be decoupled so that machine learning models can be trained locally with local data (see \S\ref{sec:ml-workflow}). 
Consequently, the availability and quality of local data are vital for the security of the edge machine learning system. 
Unlike threats, vulnerabilities, and exploits that have been analyzed in \S\ref{sec:ml-workflow} and \S\ref{sec:general-data-vul}, we discuss security issues of edge-ML data collection and curation procedures.

\begin{enumerate}
    \item \textit{Data quality}. The quality of edge-ML data plays an important role in training edge-ML models. 
    Many prior works~\cite{okafor2020improving,mahdavinejad2018machine} have found that poor data quality can dramatically degrade the performance of edge-ML models.
    To undermine edge-ML systems, data quality attacks~\cite{bostami2019false,huang2020adversarial} aim to decrease the quality of the collected data from edge devices by various approaches.
    For example, genetic attacks and probability-weighted packet saliency attacks have been found to be used to compromise edge intrusion detection systems, where attackers inject a large number of low-quality network packets through DDoS attacks~\cite{huang2020adversarial}.
    \item \textit{Data availability}. The local data have to be available for local training and evaluation to train edge-ML models in a decentralized way.
    Data availability attacks~\cite{qaim2018draw,chanal2020security} are the new threats to edge-ML systems, where attackers compromise edge device sensors to impede data collection and curation.
    For example, onboard sensors are used in autonomous driving systems to collect location data, and attackers can perform electromagnetic pulse attacks to damage electronic sensors and make the data unavailable for training autonomous driving models~\cite{petit2014potential}.   \looseness=-1
\end{enumerate}
\section{Privacy}
Since a large amount of sensitive data is uploaded from edge devices to numerous offsite servers, it is essential to ensure that user's privacy is protected from outside threats. 

\subsection{Privacy Threats}
As the data is transferred, processed, and stored on the server, it is susceptible to exploits from both an outside adversary attacking the system and network and an inside adversary sniffing information from data on the server.
Such exploits can include linkability, identifiability, exposure, and policy non-compliance. 

\subsection{User Privacy}
We separate potential vulnerabilities and exploits against users' privacy into two categories: location privacy and data privacy. The former concentrates on privacy during communications between edge device and edge server, while the latter concentrates on data within edge servers.

\PP{Location privacy}
Location privacy includes preventing the user's location information from tracking and profiling attacks. 
\begin{enumerate}
    \item \PPi{Location tracking}
        When an edge device (such as smartwatches or other mobile devices) makes a request to an edge server, 
        its location information and timestamp can be included as the metadata of the sent message or as a direct functionality \cite{tdkif2017mobile}. An untrusted server or network attacker can extract and record location data from the aforementioned device and map out the user's movements over time \cite{gg2003anonymous}.
    \item \PPi{Location profiling}
        Besides activity tracking, edge device location can also be used to establish a user profile and further help narrow down or pinpoint the user's identity. An attacker can analyze the device's most-frequent geographic coordinates and infer the user's neighborhood or workplace. Such auxiliary information can then be applied to an anonymized dataset to separate entries that belong to the user with high probability.
\end{enumerate}

\PP{Data privacy}
Data privacy focuses on protecting the user's personally identifiable information from attacks such as linking, exposing, and de-anonymizing the offloaded data. 
\begin{enumerate}
    \item \PPi{Data leakage and exposure}
        As discussed in section \ref{sec:general-data-vul}, user data in edge servers are vulnerable to thievery or leaked through various methods. If privacy policies and data anonymization techniques are not in place, the exposed data may iclude the user's name, address, and contact information, among others. 
    \item \PPi{Data linkability}
        With an anonymized dataset (e.g. with the k-anonymity model \cite{sweeney2002kanonymity}), attackers cannot directly identify users from decrypted data. However, attacks can still be carried out on such datasets \cite{sylx2014inference}. An edge device may upload the same user's data to different edge servers, resulting in the user's record existing across several datasets. An adversary can perform a links attack by analyzing the overlapping entries and identifying records that correspond to the same individual. 
    \item \PPi{Data de-anonymization}
        Due to the increased deployment of large edge networks in the age of ``big data", there is an abundance of datasets to cross-reference against each other. 
        With the help of publicly available information, datasets stored on edge servers can also be de-anonymized via re-identification attacks. Such attacks can be carried out on privacy-insensitive data \cite{alam2021person} or privacy-preserving data analytics \cite{shcll2020reidentification} that are common practices in medical and business fields. 
\end{enumerate}

\subsection{Privacy Policies}
Companies use Privacy policies to disclose how users' data are gathered, used, managed, and disclosed. These legal documents are usually extremely vague, lengthy, and complicated to read, which can cause users to skip over them and give unaware consent \cite{theatlantic-privacypolicy,cnbc-privacypolicy}. In the premises of edge computing, new threats and vulnerabilities emerge as user data is distributed among edge devices and servers in a vast network. We briefly describe some aspects in which user privacy may be violated in the edge network due to non-compliance and give further explanation and analysis in \ref{sec:compliance}

\PP{Lack of enforcement}
Without a proper non-compliance detection mechanism on each edge node, a rogue server can store, process, or disclose unauthorized user data or fail to anonymize data before transmission.

\PP{Policy conflicts}
Edge devices and servers are deployed in countless locations across the globe, where each country or organization has its own policy requirements. As data is transmitted across the network, it can move across continents and arrive at a destination with policies conflicting with its origin. For example, certain information may be considered publicly accessible, or the duration of data storage may have a longer time frame.
\section{User and Identity and Access Management}

Identity and Access Management (IAM)
ensures that the same identity
is managed for all service interactions
while simultaneously ensuring security.
It is used to authenticate an entity and
grant or deny accesses to data and
other system resources.

Usually,
a large-scale system or service
does not maintain its own identity store
or authentication mechanism to authenticate.
%
%
IAM makes the identity management
simpler for large-scale distributed systems.
%
%
It mainly 
deals with identifying entities
and managing access to resources
based on pre-established policies~\cite{iam1}.
%
Many organizations
offer IAM systems,
including SailPoint~\cite{sailpoint}, IBM~\cite{ibm-iam}, Oracle~\cite{oracle}, RSA~\cite{rsa}, and Core security~\cite{core}.

There are a number of components
related to IAM~\cite{iam2}, including
1) identity management and provisioning,
2) authentication management,
3) federated identity management, and
4) authorization management.
Those components collaboratively ensure that
authorized users are securely and effectively
incorporated into the cloud.
We will explore how each of those areas
might be affected in the setting of edge computing.

\begin{enumerate} 
\item \PPi{Diversity of identity information and APIs in edge computing}
Identity management is highly related to
the security issues of identity management in
edge computing.
Due to the diversity of devices,
it is challenging to properly
collaborate with all different types of interfaces,
including proprietary ones.
Competition among major vendors
further poses obstacles for a uniformed
management system.




\item \PPi{Forgeable identity of edge devices}
As more devices are out in the field
and provide more direct public access,
verification of submitted identity information 
becomes critical to correctly enforce further operations
of the IAM system.
For example,
preventing identity spoofing becomes
much more challenging with edge computing,
where a mechanism to construct
unforgeable identity characteristics is under urgent call.
However,
unlike users and general-purpose computers
that share a more similar set of identity characteristics
(e.g., passwords and fingerprints for human beings,
motherboards and OS versions for general-purpose computers),
each edge device embeds unique identity characteristics
depending on the vendor and the hardware,
including customized and proprietary components.
It requires tremendous efforts from
both the manufacturer and the services provider to
build a chain of trust together for edge device identities.



\item \PPi{Rogue identity providers in edge computing}
Just like the threats from rogue
Certificate Authorities (CAs) in PKIs,
similar problems exist in identity providers.
Delegating the identity management task
to a trusted third party also means that 
authenticity of the identity
and all of the credentials are
in others' hands,
and one can only hope the provider
remains trusted and benign.
In the case of rogue identity providers,
the chain of identity will completely break
under IAM.
Such a threat will even be more daunting and influential
under edge computing,
if there is any identity service for edge devices,
due to the uncountable volume of entities that will get affected.
%



\item \PPi{Lack of suitable access control models for numerous heterogeneous devices}
The access control models
designed for general-purpose computers
and cloud computing
follow a more uniform pattern
thanks to the consistency of
architecture and form factors.
%
%
However,
no general model would
be satisfying
in a complex edge computing infrastructure
due to the diversified systems
and their enabled applications,
which calls for a fine-grained access control
that can handle countless scenarios.
Not to mention,
exhaustively enumerating and making sense
of all possible threats itself
is an open research question.



\item \PPi{Loose network access control}
Often, organizations adopt threat models with weak adversarial assumptions. For instance, they may assume that the adversary may not be able to compromise end devices. Therefore, access control is very loosely defined within the network while enforcing more access control on network perimeters (i.e., firewalls). However, this becomes a major issue once the presumably trusted devices misbehave. For example, a compromised IoT device may start probing resources on the network or try to spread the infection further in the network. Such an attack is feasible since access control is non-existent or very loosely defined~\cite{ronen2017iot}. In other words, one needs to precisely define who can access what on the network and for what reason~\cite{katsis2021can, katsis2022neutron}. 
\end{enumerate}

\section{Regulatory Compliance}
\label{sec:compliance}
The emergence of edge computing and other data-driven technologies (IoT, big data, and cloud platform services) sparked initiatives to regulate and protect end-user cyber rights. A vast amount of data are produced by end-users and devices (e.g., cameras, sensors, and smartwatches) and transmitted/processed over the edge network. These data include personally identifiable information, user actions, habits, health information, etc. In edge computing, these data are transmitted to edge devices for fast processing and response, but they might as well be sent further into the network (edge servers, cloud). The network nodes could be geographically spread or even on different continents. Different privacy regulations may apply depending on where the data is transmitted, processed, and stored. For instance, European countries have adopted the General Data Protection Regulation (GDPR)~\cite{gdpr}, while the state of California in the United States has enforced the California Consumer Privacy Act (CCPA)~\cite{ccpa}. Other regulations exist to protect sensitive data, such as healthcare data (HIPAA)~\cite{hipaa}.

Edge computing deployments must be compliant with such regulations. And given the size and distributed nature of the infrastructure, this could be a challenge. Each edge node in the infrastructure must only receive, process, and maintain the data needed to perform its operations successfully (aka \textit{data minimization}). In addition, data might need to be erased or kept for a specific amount of time, depending on the effective regulation. Also, data anonymization may need to be applied before transmitting data from one place to another.

One must ensure that the edge infrastructure and operations comply with any enforced regulations. Considering the size and the geographical distribution of nodes, this could be very challenging:
\begin{enumerate}
    \item \PPi{Regulation Applicatbility} Administrators need to identify regulations that pertain to the type of data processed in each edge computing node.  
    \item \PPi{Compliance of technical operations} It requires comparing the regulatory requirements with technical device operations. Such operations need to be identified systematically to evaluate the compliance accurately and fairly across the edge network.
    \item \PPi{Compatibility issues} Regulations may pose compatibility problems across the edge network, such as the scenario in which two edge devices are in different geographical locations where different laws apply, and both devices have the same purpose. Also, assume that the processing requires access to historical healthcare data. The regulation applying to one of the edge devices is GDPR, while the other device is HIPAA. GDPR protects the right to be forgotten and disallows the storage of historical information. HIPAA does not grant the right to be forgotten; hence historical data are available on the device.
    This is a simple scenario that raises compatibility issues. Although the edge devices have the same purpose, their underlying functionality changes as one device has access to historical data while the other is not.
\end{enumerate}

Ensuring compliance on edge is not trivial. Edge network administrators need to be equipped with tools to aid the compliance process or maintain compliance. We need a systematic approach for formalizing each device's functionality on the edge network. Then we need to map the functionality to the regulatory requirements affecting the different subnetworks of the edge network and pinpoint where the issues are and how to fix them.


\section{Conclusions \& Future Work}

In this paper, we have studied edge computing security from multiple points of view spanning from the hardware, cryptography, and network to machine learning, data, and compliance. We argue that the threats and vulnerabilities emerging from each of those stacks can downgrade infrastructure security to unimaginable levels. Thus, when dealing with the edge infrastructure security, it is our position that one must take a holistic approach. 

The configuration and functionality of the edge network, the communication protocols used, and the cloud components must be well-understood and thoroughly examined. Such an analysis will reveal the soft spots in the network, which attackers may leverage to penetrate the network. The level of criticality related to vulnerabilities introduced by soft spots changes based on factors such as risk, exploitability, and impact. Of course, due to the complexity of real edge networks, a holistic analysis cannot be carried out manually. Thus, we plan to design and implement a system that will perform a holistic security analysis of a given edge network infrastructure.

\bibliographystyle{abbrv}
\bibliography{sample-base}


\end{document}